\documentclass[twosided]{article}

\usepackage[top=2cm, bottom=2cm, left=1.4cm, right=1.7cm, paperwidth=16.2cm, paperheight=24.2cm]{geometry}
\usepackage{amsmath}
\usepackage{amsthm}
\usepackage{fancybox}
\usepackage{bbm}
\usepackage{pdflscape}
\usepackage{algorithm}
\usepackage{algpseudocode}
\usepackage{caption}
\usepackage{subcaption}
\usepackage{placeins}
\usepackage{array}
\usepackage{multirow}
\usepackage{marvosym}
\usepackage{longtable}
\usepackage[version=3]{mhchem}
\usepackage{natbib}
\usepackage{textcomp}
\usepackage{rotating}
\usepackage{enumitem}
\usepackage{pifont}
\usepackage{tikz}
\usepackage{hyperref}

\usepackage{lmodern}

\usepackage{scrextend}
\KOMAoption{fontsize}{9pt}

\usepackage{caption}
\usetikzlibrary{shapes.geometric, arrows}

\tikzstyle{forwardbox} = [rectangle, rounded corners, text centered, draw=black, fill=red!30]
\tikzstyle{backwardbox} = [rectangle, rounded corners, text centered, draw=black, fill=green!30]
\tikzstyle{endbox} = [rectangle, text centered, draw=white]
\tikzstyle{arrow} = [thick,->,>=stealth]

\newcommand{\vect}[1]{\boldsymbol{#1}}

\begin{document}
\author{Ryan Ferguson\footnote{Contact: ryan.ferguson@scotiabank.com. Ryan Ferguson is a Managing Director at Scotiabank\texttrademark\, in Toronto.} and Andrew Green\footnote{Contact: andrew.green2@scotiabank.com. Andrew Green is a Managing Director at Scotiabank\texttrademark\, in London.}\footnote{The views expressed in this article are the personal views of the authors and do not necessarily reflect those of Scotiabank\texttrademark. \texttrademark\, Trademark of The Bank of Nova Scotia and used under license. Important legal information and additional information on the trademark may be accessed here: \url{http://www.gbm.scotiabank.com/LegalNotices.htm}}}

\title{Deeply Learning Derivatives}
\date{14/10/2018\vskip5mm Version 2.1}

\maketitle

\begin{abstract}
 This paper uses \emph{deep learning} to value derivatives. The approach is broadly applicable, and we use a call option on a basket of stocks as an example. We show that the deep learning model is accurate and very fast, capable of producing valuations a million times faster than traditional models. We develop a methodology to randomly generate appropriate training data and explore the impact of several parameters including layer width and depth, training data quality and quantity on model speed and accuracy.
\end{abstract}

\section{Introduction}\label{sec:introduction}

\subsection{The need for speed}


Quantitative Finance is a demanding taskmaster. Never satisfied with the progress of Moore's Law, a burgeoning cadre of quantitative analysts is employed to find new techniques that deliver more accuracy with lowered computational effort. We have a long list of changes we would make to our models if we had the computational resources. Computationally burdensome valuation adjustments are becoming increasingly important elements of derivative valuation and pricing\citep[see for example][]{Green2015b}. New regulations like the Fundamental Review of the Trading Book, add further computational costs.

One approach to meeting these challenges is the development of approximations to computationally expensive valuation functions. These techniques are commonly applied in XVA models, where trades are typically valued repeatedly inside a Monte Carlo simulation. Techniques from \emph{Approximation Theory} such as Chebyshev interpolation techniques have been applied in the context of XVA and more generally \citep{Zeron2017a,Gass2015a}.

Deep Neural Networks (DNNs) bring major benefits to the approximation of functions:
\begin{list}{}{}
	\item[$\bullet$] The \emph{Universal Approximation theorem} states that simple feed forward networks can represent a wide variety of functions under mild assumptions about the activation function \citep{Cybenko1989a, Hornik1991a}
	\item[$\bullet$] DNNs  bifurcate accuracy and valuation time. The catch is that computation time must be invested upfront in training the neural networks. 
	\item[$\bullet$] DNNs do not suffer from the curse of dimensionality. The techniques developed in the paper can be applied to valuation models with hundreds of input parameters.
	\item[$\bullet$] DNNs are broadly applicable. They can be trained by a wide variety of traditional models that employ Monte Carlo simulation, finite differences, binomial trees, etc. as their underlying framework.
	\end{list}

\subsection{Neural Networks in Finance}
Neural networks are perhaps best known in finance in the context of predictive algorithms for use as trading strategies \citep[see for example][]{Tenti1996a}. Credit scoring and bankruptcy prediction also have also seen application of neural network models. Neural networks have recently been applied to CVA, alongside other machine learning techniques, where a classifier approach has been used to map CDS to illiquid counterparties \citep[][]{Brummelhuis2017a}. Neural networks previously have been applied to the approximation of derivative valuations. \citet{Hutchinson1994a} applied neural networks to the Black-Scholes model in an early financial application. More recently, \citet{Culkin2017a} applied Deep Neural Networks to the same problem. Recently the development of the \emph{Deep BSDE solver} offers a new way of solving Backward Stochastic Differential Equations in high dimension \citep{E2017a,Labordere2017a}. BSDE have applications in finance, stochastic control and elsewhere
\subsection{Applying Deep Learning to Derivatives Valuation}
This paper makes the following contributions:
\begin{enumerate}
\item Demonstrates the use of deep neural network models as approximations to derivative valuation routines and provides a basket option as an example.
\item Develops a training methodology for neural network models, where the training and test set are generated using Monte Carlo simulation.
\item Shows that the final deep neural network has substantially lower error that the random error of the Monte Carlo models used to train it. The neural network learns to remove random Monte Carlo noise.
\item Explores a range of geometries for neural network models.
\item Provides an assessment of computational performance, while making use of CPU and GPU parallelism during training set generation and network parameter fitting.
\end{enumerate}

The remainder of this paper is organised as follows. In Section \ref{sec:DNN}, we provide a brief overview of deep neural network models and the associated fitting algorithm. In Section \ref{sec:DNNDeriv} we describe the application of deep neural networks to valuing derivatives by using a call option on a basket of stocks as an example. The example is used to demonstrate the accuracy and performance of various deep learning models. Section \ref{sec:conclusions} concludes with applications and future work. 

\section{Deep Neural Networks}\label{sec:DNN}

\subsection{Introducing Neural Networks}

\begin{table}
	\centering
	\begin{tabular}{|p{3cm}|p{9cm}|}\hline
		{\bf Notation} & {\bf Description}\\\hline
		$\vect{x}; x_j$ & input layer, jth element of input layer\\
		$\vect{y}$ & Output (vector or scalar depending on problem context)\\
		$\hat{y}^{(i)}$ & Output value from the neural network for training example $i$\\
		$(\vect{x}^{(i)}, \vect{y}^{(i)})$ & ith training example\\
		$(\vect{X}, \vect{Y})$ & Matrices of all training examples\\
		$\vect{W}^{[l]}; W_j^{[l]}$ & Weight matrix for the lth layer and jth component vector\\
		$\vect{b}^{[l]}; b_j^{[l]}$ & Bias vector for the lth layer and jth element\\
		$L$ & Number of layers in the neural network\\
		$g(z)$ & Non-linear activation function\\
		$\vect{Z}^{[l]} = \vect{W}^{[l]} X + \vect{b}^{[l]}$ & Result of matrix operations on layer l\\
		$\vect{a}^{[l]} = g(\vect{Z}^{[l]})$  & Results of activation function on layer l\\
		$\vect{a}^{[0]} = \vect{x}; \vect{a}^{[L]} = \vect{y}$& Input and output layers\\
		$J(\vect{W}^{[l]}, \vect{b}^{[l]})$ & Cost function\\
		$\odot$ & Element-wise multiplication\\\hline
	\end{tabular}
\caption{\label{tab:notation}Neural network notation used in this article.}
\end{table}

An artificial neural network consists of a series of layers of \emph{artificial neurons}.\footnote{For a detailed introduction to Deep Learning, see \citet{Goodfellow2016a} and \citet{Ng2018a}.}\footnote{A summary of the notation used in this article can be found in Table \ref{tab:notation}.} Each neuron takes a vector of inputs from the previous layer, $a^{[l-1]}_j$ and applies a weight vector, $W^{[l]_j}$ and a bias $b^{[l]}_j$ so that  
\begin{equation}
Z^{[l]}_j = W^{[l]}_j a^{[l-1]} + b^{[l]}_j
\end{equation}
or in matrix notation
\begin{equation}\label{eq:zdef}
\vect{Z}^{[l]} = \vect{W}^{[l]} \vect{a}^{[l-1]} +\vect{b}^{[l]}.
\end{equation}
The result from layer $l$ is then given by applying a non-linear \emph{activation function} $g(Z)$,
\begin{equation}\label{eq:adef}
\vect{a}^{[l]} = g(Z^{[l]}). 
\end{equation}
A number of different activation functions can be used and the main types are listed in Table \ref{tab:actfunc}. In order to produce non-linear outputs it is necessary that a non-linear function be used. All neurons in the same layer use the same activation function but different layers often use different activation functions. Equations \eqref{eq:zdef} and \eqref{eq:adef} together describe how the input $\vect{a}^{[0]} = \vect{x}$ vector is \emph{forward propagated} through the network to give the final output $\vect{a}^{[L]} = \vect{y}$.

\begin{table}
	\centering
	\begin{tabular}{|l|c|}\hline
		{\bf Function} & {\bf Definition}\\\hline
		Sigmoid & $ 1 / (1 + e^{-z})$ \\
		tanh & $(e^z - e^{-z}) / (e^z + e^{-z})$ \\
		ReLU & $\max(0, z)$\\
		Leaky ReLU & $\max(0.01z, z)$\\\hline
	\end{tabular}
\caption{\label{tab:actfunc} The four main types of activation function defined}
\end{table}

\begin{figure}[htpb]
	\centering
	\includegraphics[width=\textwidth]{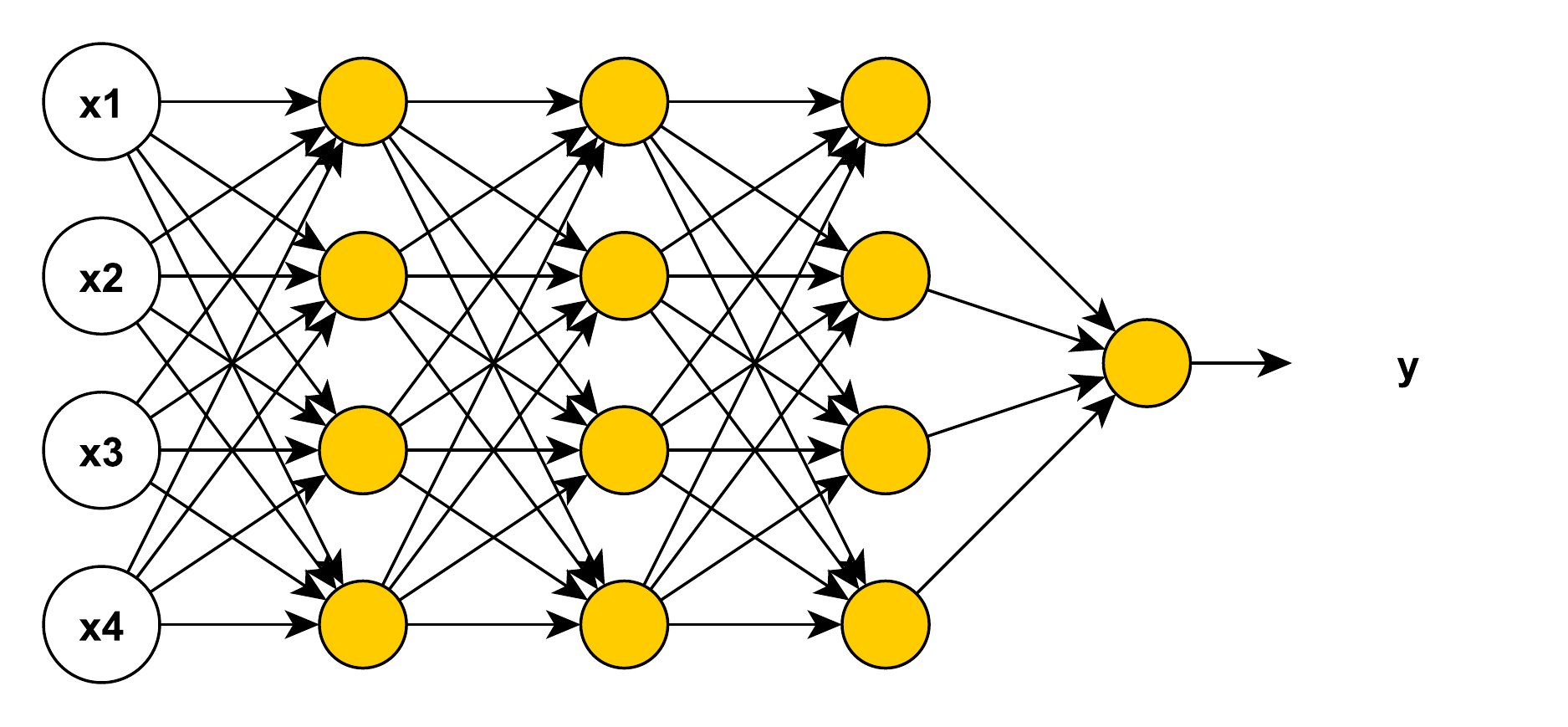}
	\caption{\label{fig:dnn1}A Neural Network with four inputs, one output and three hidden layers.}
\end{figure}

The number of inputs to the model is dictated by the number of input \emph{features}, while the number of neurons in the output layer is determined by the problem. For a regression problem with one real-valued output, as described in this paper, there will be a single node in the output layer. If a neural network model is used in a classifier problem with multiple classes, there will be one node per class. The number of layers in the model, $L$ and the number of neurons in each layer can be considered \emph{hyperparameters} and these are normally set using hyperparameter tuning over a portion of the training data set, as described in Section \ref{sec:hypertuning}. Models with many hidden layers are known as \emph{deep neural networks}. A example neural network geometry is illustrated in Figure \ref{fig:dnn1}.

\subsection{Training the Model and Back Propagation}

During the training phase, the weights and bias parameters of the model are systematically updated to minimize the error between the training data and estimates generated by the model. A set of $m_{\text{train}}$ training examples is used and the error is given by a cost function $J$,
\begin{equation}
J(\vect{W}^{[l]}, \vect{b}^{[l]}) = \sum_i^{m_{\text{train}}} \mathcal{L}(\hat{y}^{(i)}, y^{(i)} ) = \vect{\mathcal{L}}(\vect{\hat{y}}, \vect{y} ),
\end{equation}
where the function $\mathcal{L}$, depends on the choice of loss measure. A number of choices are available for the loss measure including the L2 or L1 norm.

The minimum is found using an optimization procedure. The standard approach uses batch gradient descent or a variant such as mini-batch gradient descent.\footnote{In this paper we use mini-batch gradient descent where the size of each mini-batch is a hyperparameter.} Such optimization procedures require the gradients of the cost function,
\begin{eqnarray}
d\vect{W}^{[l]} = & \frac{\partial J}{\partial \vect{W}^{[l]}}\\
d \vect{b}^{[l]} = &  \frac{\partial J}{\partial \vect{b}^{[l]}},
\end{eqnarray}
which are used to update the weights,
\begin{eqnarray}
\vect{W}^{[l]} = & \vect{W}^{[l]}  - \alpha d\vect{W}^{[l]} \\
\vect{b}^{[l]} = & \vect{b}^{[l]}  - \alpha d\vect{b}^{[l]}. 
\end{eqnarray}

\begin{figure}
	\centering
	\begin{tikzpicture}[node distance=1.5cm]
	
	\node (inputforward) [endbox] {$a^{[0]}$};
	\node (fwd1) [forwardbox, right of=inputforward] {$W^{[1]}, b^{[1]}$};
	\node (fwda1) [endbox, right of=fwd1] {$a^{[1]}$};
	\node (fwddots) [endbox, right of=fwda1] {$\dots$};
	\node (fwdL) [forwardbox, right of=fwddots] {$W^{[L]}, b^{[L]}$};
	\node (fwdaL) [endbox, right of=fwdL] {$a^{[L]} = \hat{y}$};
	\node (fwdZ1) [endbox, below of=fwd1] {$Z^{[1]}$};
	\node (fwdZL) [endbox, below of=fwdL] {$Z^{[L]}$};
	\node (bkwd1) [backwardbox, below of=fwdZ1] {$dZ^{[1]}$};
	\node (bkwda1) [endbox, right of=bkwd1] {$da^{[1]}$};
	\node (bkwddots) [endbox, right of=bkwda1] {$\dots$};
	\node (bkwdL) [backwardbox, right of=bkwddots] {$dZ^{[L]}$};
	\node (bkwdaL) [endbox, right of=bkwdL] {$da^{[L]}$};
	\node (bkwddWdb1) [endbox, below of=bkwd1] {$dW^{[1]}, db^{[1]}$};
	\node (bkwddWdbL) [endbox, below of=bkwdL] {$dW^{[L]}, db^{[L]}$};
	
	\draw [arrow] (inputforward) -- (fwd1);
	\draw [arrow] (fwd1) -- (fwda1);
	\draw [arrow] (fwda1) -- (fwddots);
	\draw [arrow] (fwddots) -- (fwdL);
	\draw [arrow] (fwdL) -- (fwdaL);
	\draw [arrow] (fwd1) -- (fwdZ1);
	\draw [arrow] (fwdL) -- (fwdZL);
	\draw [arrow] (fwdZ1) -- (bkwd1);
	\draw [arrow] (fwdZL) -- (bkwdL);
	\draw [arrow] (bkwdaL) -- (bkwdL);
	\draw [arrow] (bkwdL) -- (bkwddots);
	\draw [arrow] (bkwddots) -- (bkwda1);
	\draw [arrow] (bkwda1) -- (bkwd1);
	\draw [arrow] (bkwd1) -- (bkwddWdb1);
	\draw [arrow] (bkwdL) -- (bkwddWdbL);

	\end{tikzpicture}
	\caption{\label{fig:fwdbackprop}Forward and backward propagation algorithms for a Neural Network with $L$ layers (see text for full description)}
\end{figure}
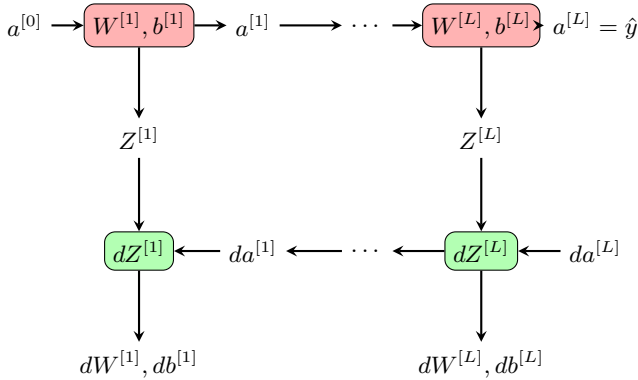

To obtain the gradients a specialised type of \emph{algorithmic differentiation} is used, known as \emph{backpropagation}. This algorithm recursively applies the chain rule to propagate derivatives back through the computational graph of the neural network model. Beginning with the output layer, the derivative $d\vect{a}^{[L]}$ is defined by
\begin{equation}
d\vect{a}^{[L]} = \nabla_{\vect{a}^{[L]}=\vect{\hat{y}}} \vect{\mathcal{L}}(\vect{\hat{y}}, \vect{y} ).
\end{equation}
The derivative $d\vect{Z}^{[L]}$ is then given by,
\begin{equation}
d\vect{Z}^{[L]} = d\vect{a}^{[L]} \odot g^\prime(\vect{Z}^{[L]}).
\end{equation}
Hence the derivatives of the weights for layer L are given by,
\begin{eqnarray}
d\vect{W}^{[L]} = & \frac{1}{m} d\vect{Z}^{[L]} . \vect{a}^{[L-1]}\\
d\vect{b}^{[L]} = & \frac{1}{m} d\vect{Z}^{[L]}.
\end{eqnarray}
The derivative $d\vect{a}^{[L-1]}$ for the next layer are then given by,
\begin{equation}
d\vect{a}^{[L-1]} = (\vect{W}^{[L]})^T . d\vect{Z}^{[L]},
\end{equation}
and hence the derivatives can be propagated backwards through the network to obtain all the $d\vect{W}^{[l]}$ and $d\vect{b}^{[l]}$. Forward and backward propagation are illustrated in Figure \ref{fig:fwdbackprop}.

A set of data is used to train the model. Typically, a data set of $m$ examples is divided into three independent groups, $m_{\text{train}}~$, which is used to train the network, $m_{\text{test}}$ which is used to test the trained network and $m_{\text{dev}}$ which is used to develop or cross-validate the hyperparameters of the model. The fraction of data divided into each group depends on how much training data is available, with $m_{\text{test}}$ and $m_{\text{dev}}$ forming a relatively larger proportion if $m$ is relatively small. The training stops after a fixed number of iterations, or when the loss function is no longer decreasing. 
	
The weights are initialized using a pseudo-random number generator, with samples drawn either from the uniform distribution or the standard normal distribution. One common procedure with the ReLU activation function initializes the weights in layer $l$ using a standard normal distribution with mean zero and variance $2 / n^{[l]}$.

\subsection{Hyperparameter Tuning}\label{sec:hypertuning}
While the weights, $\vect{W}$ and the bias parameters, $\vect{b}$, are defined during training, other parameters such as the \emph{learning rate} $\alpha$ and the topology of the neural network are \emph{hyperparameters} that are not directly trained. The hyperparameters could be specified exogenously, however, in general their optimal values are not known \emph{a priori}. Hence hyperparameters are optimised by assessing model performance, once trained, on a separate independent set of samples. 

In this paper we focus on a few different axes of hyperparameter tuning by varying the number of nodes per hidden layer, the number of hidden layers, the size of each batch of training data, the total amount of data available for training, the quality of the training data. We demonstrate the impact these hyperparameters have on accuracy and performance.

\section{Derivatives Valuation and Deep Learning}\label{sec:DNNDeriv}

A derivatives valuation model is ultimately just a function which maps inputs, consisting of market data and trade specific terms, to a single output representing the value. That function may have an known analytic form, though frequently numerical approaches including Monte Carlo simulation, binomial trees or finite difference techniques must be used. Simple European stock options can be valued with as few as five inputs, while more complex products like Bermudan swaptions have valuation functions requiring many more inputs, involving all the properties of the underlying swap and option exercise schedule. The number of parameters could be in the hundreds or thousands for such complex products. Having a large number of inputs to a neural network model is certainly feasible and many current practical applications have inputs of a similar order of magnitude. For example, an image recognition application might use images 64 x 64 pixels in size with 3 colours, with each sub-pixel in itself a model input. However, the dimensionality of the input does scale the requirements for the size training dataset. 

When training a model to approximate a derivatives valuation function, one of the first decisions is the domain of application. We can choose to train on all the parameters of the valuation model or only some of them. Even then, we can opt to train over a large domain for a given parameter, or a much smaller domain. The trade-off is one of generality versus model-complexity and training time.  For example, in the case of a Bermudan swaption we may choose to take as given the properties of a specific trade and then only train the model against a variety of input market data scenarios. Given the reduction in size of the parameter space, the models will be smaller and their training requirements such as the size of training data set, the amount of time spent training, etc lower.

It is also important to note that the model can only be trusted to approximate the function well over the parameter ranges that were used to train it. Should inputs move outside of these ranges then it is unlikely the approximation will perform well. It is prudent to monitor the use of the model and retrain should inputs move outside the training range.

In this paper we train models which can generalize over a large number of parameters with broad ranges of applicability and leave the development of domain-specific models for future work.

\subsection{Derivatives Pricing Example: Basket Options}\label{sec:approxBasket}

To demonstrate how a network can learn a derivative valuation model we trained a network to price a European call option on a worst-of basket with six underlying stocks: 

\begin{equation}
V = \max(0,\min(Stock1,Stock2,Stock3,Stock4,Stock5,Stock6) - K)
\end{equation}

This option is commonly valued using a Monte Carlo simulation with each of the stocks following a geometric Brownian motion. This example was selected for four reasons: 
First the dimensionality of the function is moderate, with the number of input parameters equal to $\frac{1}{2}n(n+3)+1$. In our example with six underlying stocks, we have a 28-dimensional input space. 
Second, the use of Monte Carlo explores the use of deep learning with a computationally intensive valuation routine. 
Third, it will be easy to scale up the dimensionality of this problem by adding more stocks to the basket. 
Fourth, the use of a numerical method like Monte Carlo means that the function to be replicated has numerical noise.

To fit a neural network model we need to generate a set of training examples $(\vect{x}^{(i)}, y^{(i)})$, with the $\vect{x}^{(i)}$ being input model parameters and the $y^{(i)}$ being the output value. To generate the training data we use Monte Carlo simulation to generate random sets of inputs within appropriate ranges selected for each parameter. In the context of this example all parameters were generated independently with the distribution selected individually for each parameter. The choice of sampling distribution is important and should be made on a case by case basis. For example, there is little point in generating a large volume of examples which yield zero option value. Hence we anticipate that different choices of sampling distribution, where the parameters are not independent will be needed for some derivative models. In other cases, stratification or importance sampling may be appropriate but we leave this for further research.

Once a random set of parameter values is created, the function to be replicated is called to generate a value. In this way a large number of training examples can be created. This can be computationally intensive as the derivative valuation function is called many times. However, this process needs only be done once for the initial training of the model. Given all training examples are independent it can also be readily parallelized.

\subsection{Training for Basket Options}

For our basket option example, we reduced the number of input parameters by assuming flat implied volatility surfaces for each of the stocks in the basket. Further simplification was made by assuming both zero interest rates and dividends. For the case of six underlying stocks we have the parameter breakdown as shown in Table \ref{tab:basketparam}.

\begin{table}
	\centering
	\begin{tabular}{|l|c|l|}\hline
		{\bf Parameter Type} & {\bf Number} & {\bf Distribution}\\\hline
		Forward stock prices & 6 & 100(lognormal(0.5, 0.25))\\
		Stock volatilities & 6 & uniform(0,1)\\
		Maturity & 1 & uniform(1,43)**2\\
		Correlations & 15 & 2($\beta(5,2)$-0.5)\\\hline
	\end{tabular}
\caption{\label{tab:basketparam}Parameters of the basket option with six underlyings.}
\end{table}

The parameters were sampled randomly and independently, with the exception of correlation matrices which were handled separately. Stock prices were generated using $100 *\exp(z)$, with $z \sim N(0.5, 0.25)$. The correlation matrices were generated using the C-vine method of \citet{lewandowski2009a}.

\begin{figure}[htp]
	\centering
	\begin{subfigure}[b]{0.48\textwidth}
		\centering
		\includegraphics[width=\textwidth]{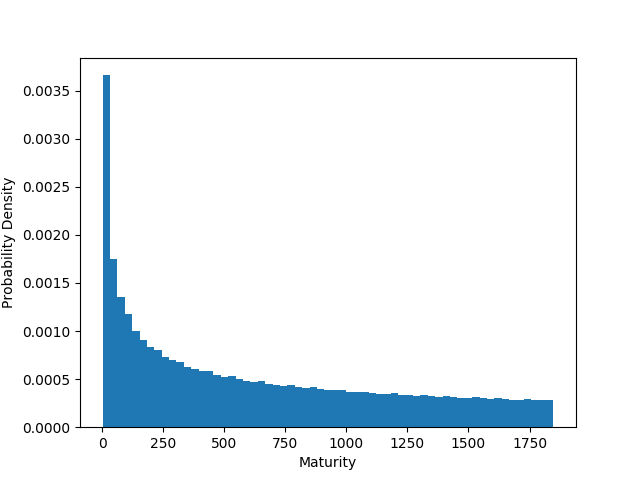}
				\caption{}
	\end{subfigure}
	\begin{subfigure}[b]{0.48\textwidth}
		\centering
		\includegraphics[width=\textwidth]{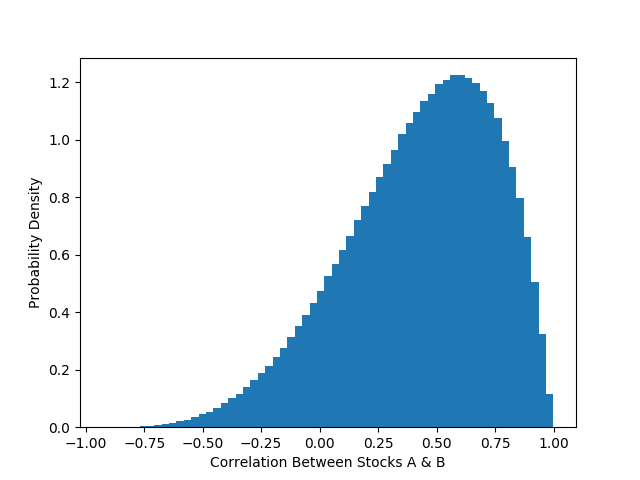}
		\caption{}
	\end{subfigure}
	\medskip
	\begin{subfigure}[b]{0.48\textwidth}
		\centering
		\includegraphics[width=\textwidth]{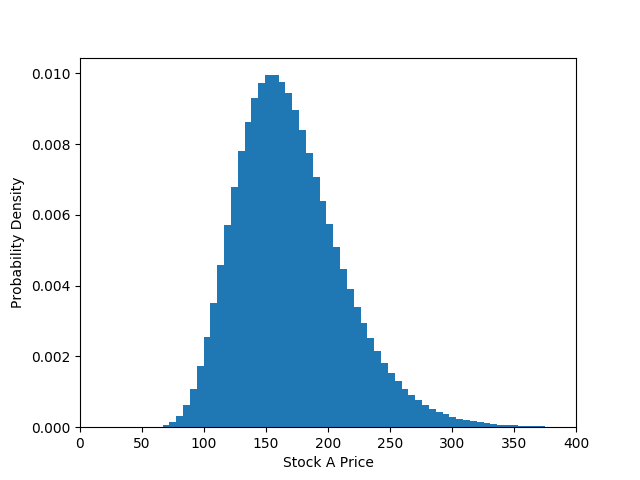}
		\caption{}
	\end{subfigure}
	\begin{subfigure}[b]{0.48\textwidth}
		\centering
		\includegraphics[width=\textwidth]{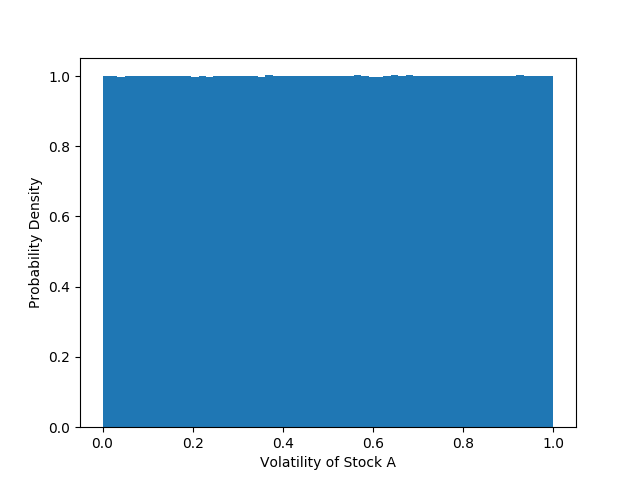}
		\caption{}
	\end{subfigure}
	\medskip
	\begin{subfigure}[b]{0.48\textwidth}
		\centering
		\includegraphics[width=\textwidth]{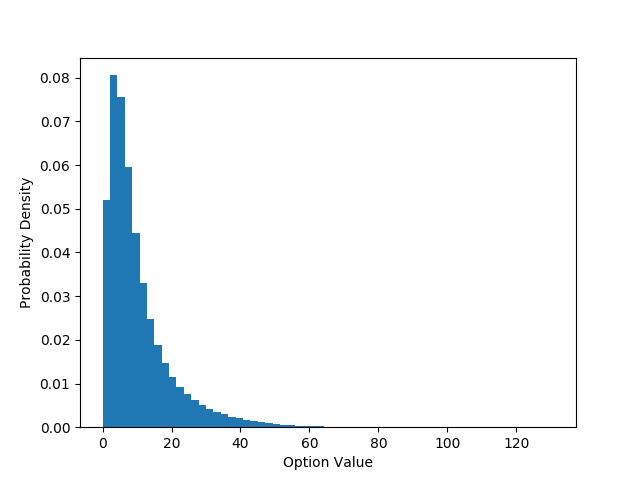}
		\caption{}
	\end{subfigure}
	\caption{\label{fig:sampledist}Sample distributions of input variables used in the basket option training and test sets. Given there a six input assets to the basket and the sampling was independent, only one representative example is illustrated for stock, volatility and correlation. The individual figures illustrate (a) maturity (in days), (b) correlation between stock A and stock B, (c) stock price A, (d) volatility of stocks and (e) option value.}
\end{figure}

Since a neural network model will minimize the error between its estimates and that of the training data that it is presented, it is important to choose a representative data set. Another consideration when developing training data is the nature of the function being learned. Areas of rapidly changing function value need to be assigned more training data. We generated more data for short maturities since the convexity of the basket option valuation function is greatest for at the money short dated options. The sample distributions of the inputs are illustrated in Figure \ref{fig:sampledist}.

We generated three different training sets using the same distributions as in Table \ref{tab:basketparam}, with properties given in Table \ref{tab:trainingsets}.  The time taken to generate each was approximately the same, approximately a week of server time using all of the cores.
\begin{table}[h]
	\centering
	\begin{tabular}{|l|c|c|}\hline
		{\bf Training Set} & {\bf \# Examples} & {\bf \# MC Paths}\\\hline
		A & 5M & 1M\\\hline
		B & 50M & 100k\\\hline
		C & 500M & 10k\\\hline
	\end{tabular}
	\caption{\label{tab:trainingsets} Properties of the three training sets.}
\end{table} 

The models were trained with minibatch sizes of 50,000 samples. The test set consisted of 5,000 samples drawn at random from a separate set of highly accurate MC generated data (100mm paths) used only for testing.  Using a 24-core AMD EPYC 7401P server we computed values for the test set in a little more than a week by fully using all of the cores. By using 100 million paths we can obtain Monte Carlo accuracy to within 1 cent. However, the valuation of a single option took approximately 300 seconds making it of little use in a production environment. If 5 cent accuracy (1 million paths) or 20 cent accuracy (100k paths) is acceptable, then valuing the test set can be completed in an hour and 45 minutes, or 10 minutes, respectively.

All models were trained using an Adam optimizer \citep{Kingma2014a} with the learning rate set to 1e-3, $\beta_1 = 0.9$ and $\beta_2 = 0.999$. 

\subsection{Results}

\subsubsection{Training Performance}

Different models, all with six hidden layers were trained. The models were differentiated by the training set used and number of nodes in each hidden layer. Each hidden layer used ReLU activation functions with the final output layer a simple linear function to give a real valued output.  

\begin{figure}[htpb]
	\centering
	\includegraphics[width=\textwidth]{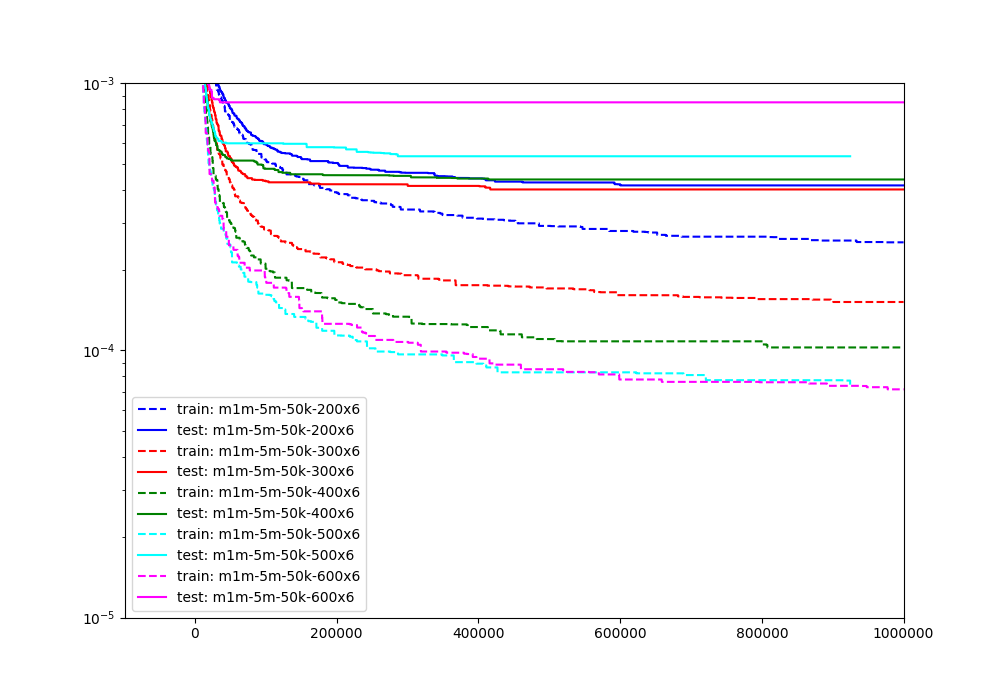}
	\caption{\label{fig:l1m}Minimum loss as a function of mini-batch iteration for five different models with 200, 300, 400, 500 and 600 nodes per layer. }
\end{figure}

The first experiment used models trained with training set A. Figure \ref{fig:l1m} illustrates the learning capacity as a function of the number of mini-batch iterations for each of the five models. Two curves are presented for each model, giving the learning curve for the test and training sets. The curves represent the minimum loss from each iteration of the Adam optimizer. This is a classic example of overtraining; the training set error continues to decrease while the test set error remains stubbornly high even for the best models. For the largest model, with 600 nodes per layer, the overfitting is so pronounced that the generalization on the out-of-sample is materially worse than that of the smaller models. 

\begin{figure}[htpb]
	\centering
	\includegraphics[width=\textwidth]{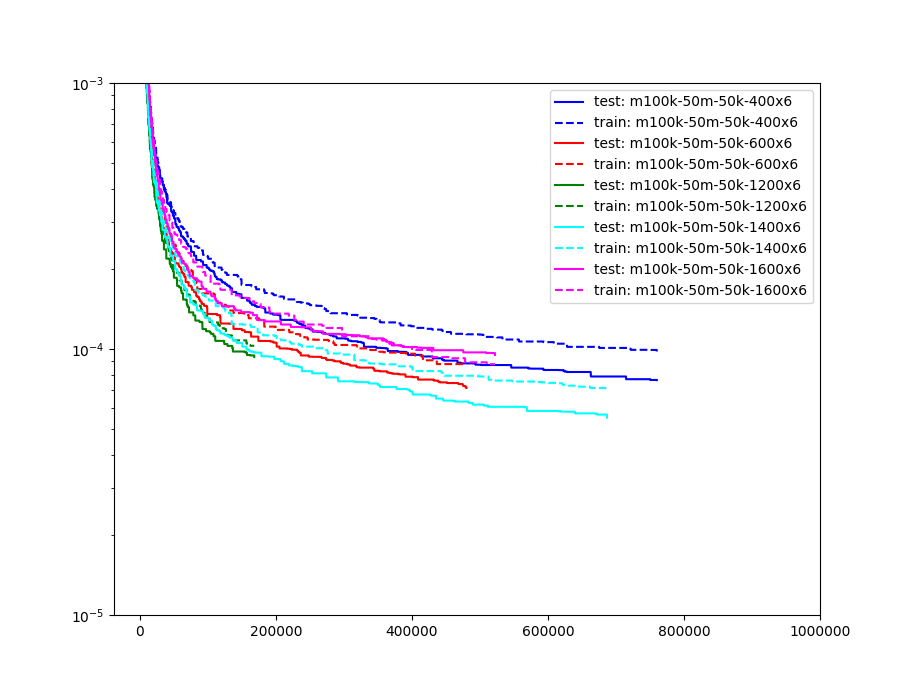}
	\caption{\label{fig:l100k}Minimum loss as a function of mini-batch iteration for five different models with 400, 600, 1200, 1400 and 1600 nodes per layer.}
\end{figure}

Training set B was used to train the second set of models. These models used 400, 600, 1200, 1400 and 1600 nodes per layer with the learning curves illustrated in figure \ref{fig:l100k}. 

\begin{figure}[htpb]
	\centering
	\includegraphics[width=\textwidth]{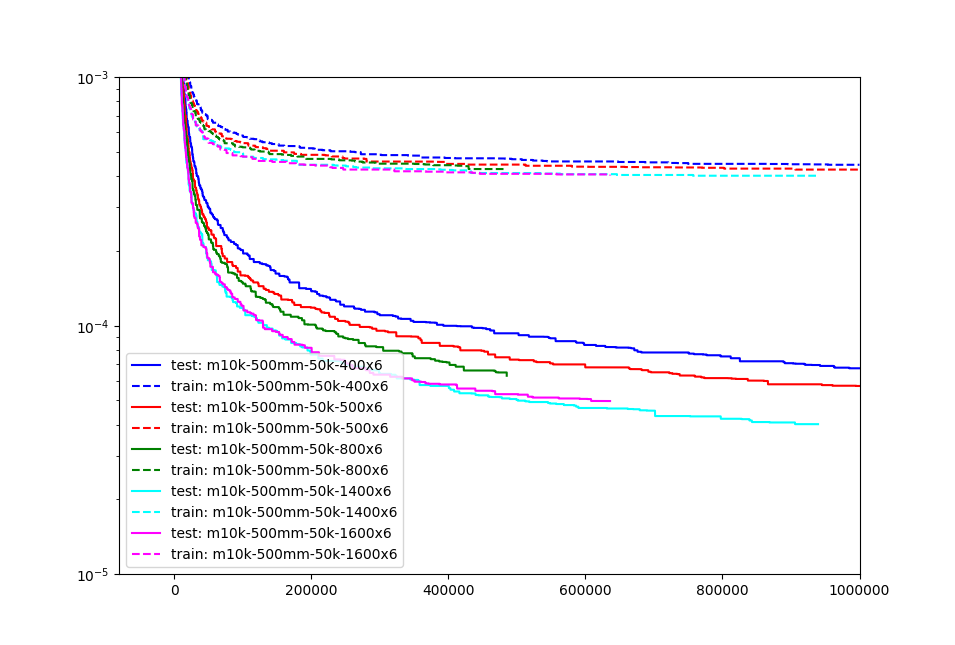}
	\caption{\label{fig:l10k}Minimum loss as a function of mini-batch iteration for five different models with 400, 600, 1200, 1400 and 1600 nodes per layer.}
\end{figure}

The final set of experiments were conducted training models with the same number of neurons per layer as case B, but using training set C, with only 10k Monte Carlo paths, and the results are illustrated in figure \ref{fig:l10k}. With a larger training set with more Monte Carlo noise in each training example, the results were significantly better. Empirically it appears better to distribute the Monte Carlo scenarios broadly through the model domain rather than to concentrate them at points generating highly accurate training data. With more data, larger models with more capacity can be trained before overfitting sets in. A key observation is that the test results are better than the training results; that is the models have learned to average out the numerical noise associated with the lower-quality training data.

The logical conclusion of this suggests that the best approach may be obtained using single Monte Carlo paths directly. Since each of our training samples are each approximately 300 bytes in size, each training set would consume 1.5 petabytes of disk space. While this is not impossibly large, training approaches using on the fly data generation seem more reasonable. We propose training Neural Networks directly from the dynamic Monte Carlo simulation, with paths grouped into independent mini-batches. However, we leave this for further work. 

\subsubsection{Accuracy}
\begin{figure}[htp]
	\centering
	\begin{subfigure}[b]{0.48\textwidth}
		\centering
		\includegraphics[width=\textwidth]{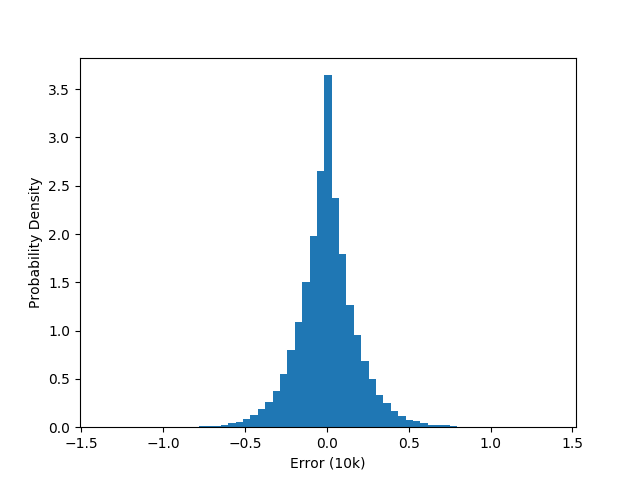}
		\caption{}
	\end{subfigure}
	\begin{subfigure}[b]{0.48\textwidth}
		\centering
		\includegraphics[width=\textwidth]{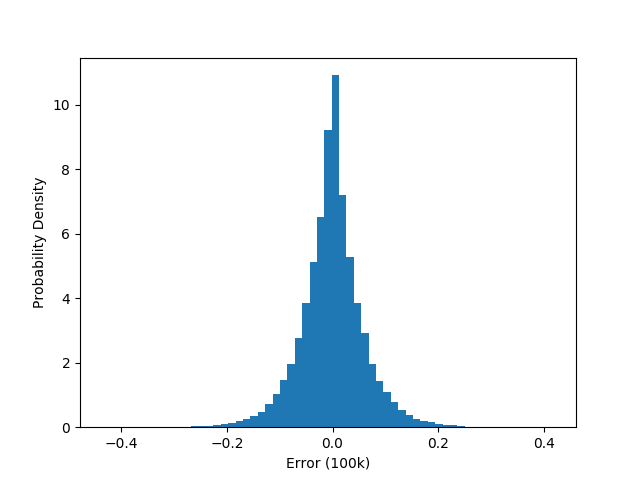}
		\caption{}
	\end{subfigure}
	\medskip
	\begin{subfigure}[b]{0.48\textwidth}
		\centering
		\includegraphics[width=\textwidth]{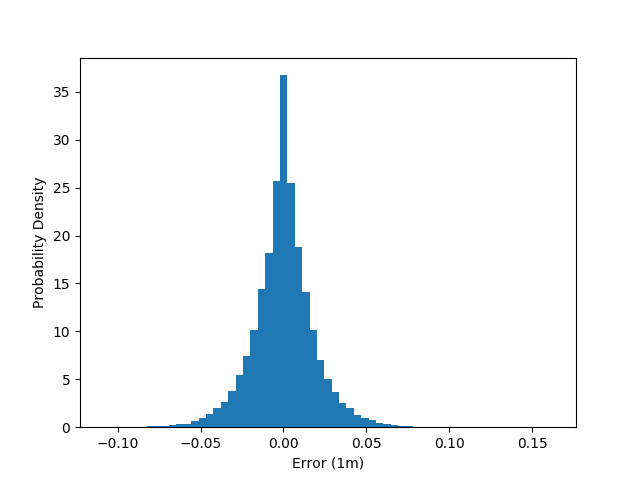}
		\caption{}
	\end{subfigure}
	\begin{subfigure}[b]{0.48\textwidth}
		\centering
		\includegraphics[width=\textwidth]{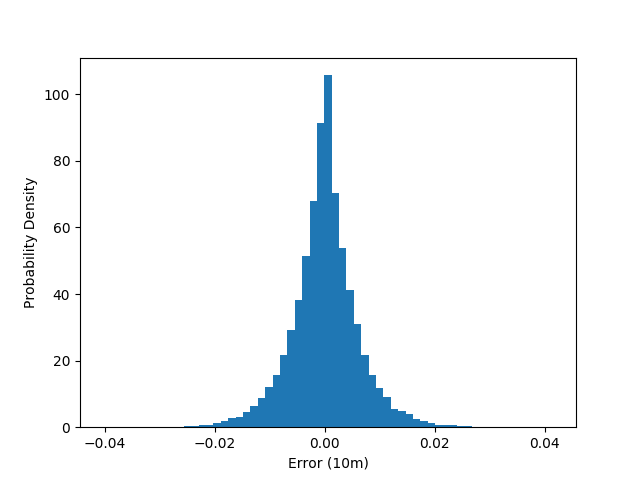}
		\caption{}
	\end{subfigure}
	\medskip
	\begin{subfigure}[b]{0.48\textwidth}
		\centering
		\includegraphics[width=\textwidth]{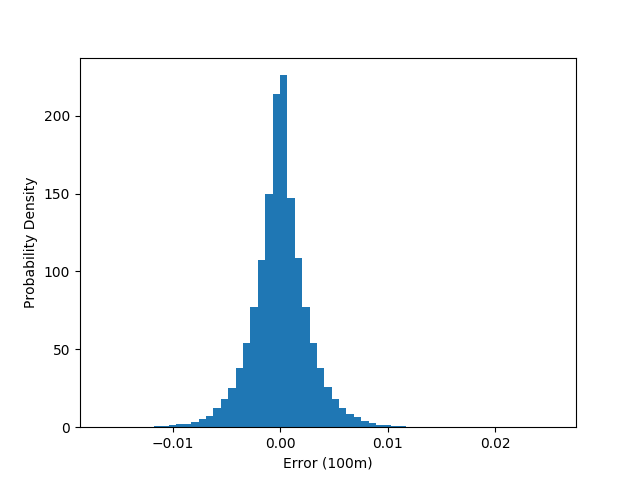}
		\caption{}
	\end{subfigure}
	\begin{subfigure}[b]{0.48\textwidth}
		\centering
		\includegraphics[width=\textwidth]{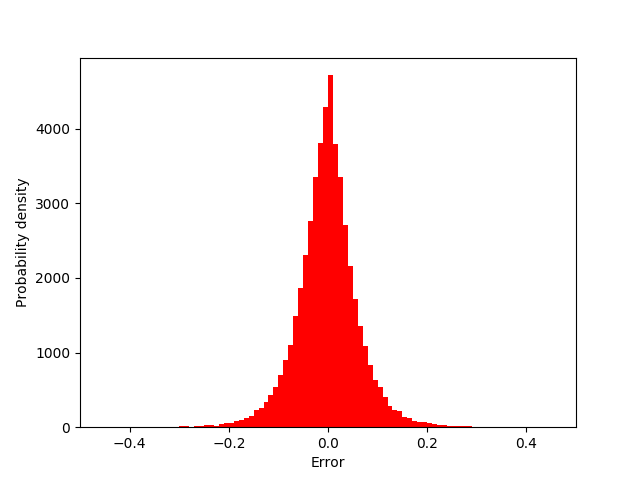}
		\caption{}
	\end{subfigure}
	\caption{\label{fig:montecarlo}Error distributions over the 50,000 sample test set evaluated by traditional Monte Carlo techniques and a deep learning model. The individual figures represent differing amounts of scenarios used in each valuation: (a) 10k, (b) 100k, (c) 1 million, (d) 10 million, (e) 100 million, and (f) Deep Learning Model. The error of the deep learning model is similar to that of the Monte Carlo simulation with 100k scenarios. The MC results with 100k scenarios took a little over 10 minutes to calculate. The deep learning model calculated its results in less than 6 milliseconds.}
\end{figure}

In order to benchmark our deep learning approach against classical Monte Carlo valuation techniques, we valued these test cases using Quantlib's MCEuropeanBasketEngine with 10k, 100k, 1mm, 10mm and 100mm scenarios. The histogram of differences compared with a second MC evaluation of 100mm paths is presented in Figure \ref{fig:montecarlo}. The figure also shows the results of one of the models trained during this study. The results are quite similar to the Monte Carlo examples with 100k scenarios. However, there is one big difference. As noted earlier, The Monte Carlo results took a little over 10 minutes to calculate using all the cores of our Epyc 7401P server. The deep learning inference results took less than 6 milliseconds to compute on an NVIDIA GTX 1080ti GPU. This was not a case of using more expensive hardware. The GPU costs less than the CPU.

We also investigated the samples in the tails of the histogram as they often came from the edges of the training distribution. Later work will investigate techniques to score the likelihood that a sample of test data will produce an accurate valuation with kernel density estimation and statistical measures of distance showing promise.

\subsubsection{Speed}
The overall performance can be broken down into the three key phases, generation of the training set, training the neural network and the final inference step. 

We generated three training datasets, trading off the size of the dataset against the quality of data (ie number of Monte Carlo scenarios per sample), while holding the total number of Monte Carlo scenarios at 5x10$^{12}$. Each dataset took approximately 1 week to generate using all 24 cores of an AMD EPYC 7401P server.

Neural network model training is a function of the number of layers, the nodes per layer, the size of the training set, the size of mini-batch and the learning rate. The models presented in this paper were trained for between 3 hours and 1 week.

The time spent generating the training set and training the model are one-off costs. Once completed they do not have to be repeated. These phases are most accurately compared to the time spent developing traditional models. Rarely do we discuss how long it took code and test our basket MC model.

The time we are truly concerned with is the inference time: how quickly can a trained model return a valuation. This varies with the size of the model. The small models we trained (6 layers with 300 nodes per layer) were capable of returning more than 20,000 valuations in parallel in less than 50 microseconds. The large models (6 layers with 1400 nodes per layer) can compute 50,000 valuations in less than 6 milliseconds.

\section{Conclusions and Opportunities for Future Work}\label{sec:conclusions}
In this paper we have demonstrated that deep neural networks can be used to provide highly accurate derivatives valuations using a basket option example. These models compute valuations approximately 1 million times faster than traditional Monte Carlo models. We have developed a successful methodology to generate a set of training data, which can be adapted to suit other valuation models. Finally we have also demonstrated that using small numbers of Monte Carlo paths in the training set is very effective and the the neural network learns to average out the random error component of the Monte Carlo model found in the training set.

We are actively investigating several areas:
\begin{enumerate}
	\item Scaling the dimensionality of the derivatives pricing model. Some models require many hundreds of inputs across market data and trade specifics.
	\item Developing scoring techniques to determine the suitability of the trained model to input data being appplied.
	\item Providing risk sensitivities through back propagation of neural networks as an alternative or complement to algorithmic differentiation.
	\item Using DNNs to approximate both valuation model and model calibration steps independently as a pipeline.
	\item Applying transfer learning to accelerate the training of related derivative products.
	\item Using DNNs in practice to accelerate derivative valuations for XVA.
	\item The capacity of a model is a function of the number of layers and the number of nodes per layer. Future work will also involve changing the depth of the models while holding the nodes per layer constant.\footnote{Further discussion of learning capacity can be found in \citet{Guss2018a}}.
	\item Exploring the impact of mini-batch size. Smaller mini-batches allow GPU memory to be used for other purposes, such as building larger models.\footnote{Early indications suggest this is not particularly important for the accuracy of the model.}
\end{enumerate}

\bibliographystyle{chicago}
\bibliography{green_book}

\end{document}